\documentstyle[11pt,newpasp,twoside,epsf]{article}
\begin{document}
\title{
A spectroscopic survey of stars brighter than B=16.6 magnitude in $\omega$
Centauri}
\author{Jacco Th. van Loon}
\affil{Astrophysics Group, School of Chemistry \& Physics,
       Keele University, Staffordshire ST5 5BG
       (jacco@astro.keele.ac.uk)}
\begin{abstract}
Using the 2dF multi-fibre instrument at the AAT, we have carried out a
spectroscopic survey of over 1500 stars brighter than B=16.6 mag, which have
been confirmed by the proper motion study of van Leeuwen et al.\ (2000) to be
members of the massive galactic globular cluster, $\omega$ Centauri. The
survey aimed at sampling the Hertzsprung-Russell Diagram in a uniform way,
thus avoiding a bias towards the dominant stellar population and allowing the
study of stars with less common properties such as particularly metal-rich or
metal-poor stars. Exceptionally, though, emphasis has been given to include as
many RR Lyrae variables as possible in order to address the question why these
variables are distributed over such a wide observed range in colours ---
coinciding with many non-variable stars of similar colours. The potential of
the data and analysis are presented together with the first preliminary
results.
\end{abstract}

\section{Introduction}

Globular clusters are unique objects, especially for the study of stellar
evolution. Each globular cluster contains $N\sim10^5$ stars, which are more or
less co-eval and at the same fairly well-known distance. This greatly
simplifies the study of the evolutionary stages after the exhaustion of
hydrogen in the core of these stars: in effect, these post-Main Sequence stars
all have about the same Zero-Age Main-Sequence (ZAMS) mass, and comprise a
relatively narrow range in metallicity. The great number of such stars,
thousands per cluster, makes it possible to accurately derive the mean and
shape of the distribution over a particular stellar parameter for a particular
evolutionary phase. Examples of short-lived phenomena may be encountered, that
last for only ${\Delta}t\sim10^5$ years. With typical distances of several
kpc, the local galactic globular clusters are sufficiently nearby that
photometric and spectroscopic studies are possible for most of the post-Main
Sequence evolutionary stages. Different globular clusters show differences in
mean metallicity that span more than two orders of magnitude between
[Fe/H]$\sim-2$ and [Fe/H]$\sim0$, whilst their ages are all very similar with
$t\sim10^{10}$ years. This allows the study of the effects of metallicity on
post-Main Sequence evolution, but at the same time it limits such studies to
stars with masses $M_{\rm ZAMS}\sim0.8$ M$_\odot$.

The most massive galactic globular cluster, $\omega$ Cen is favourably located
at a distance of $d_\odot\sim5$ kpc from the Sun and $d_{\rm GC}\sim6$ kpc
from the galactic centre (Harris 1996). Membership has been determined on the
basis of a recent proper motion survey for all stars with $B<16.6$ mag (van
Leeuwen et al.\ 2000), resulting in a near-complete sample of stars that have
evolved beyond the base of the first-ascent red giant branch (RGB). The mean
metallicity of stars in $\omega$ Cen is with [Fe/H]$=-1.6$ very typical, but a
rather large spread in metallicity has been suggested in recent years, which
has subsequently been interpreted in terms of a spread in age over a few Gyr
(Norris et al.\ 1996). As such, $\omega$ Cen is an excellent object for the
study of post-Main Sequence stars albeit with the possible complication of a
spread in abundances --- which in itself merits further investigation. Many
clusters differ from each other also in other respects, for instance the
specific frequency of X-ray binaries, and the morphology of the horizontal
branch is known to depend not only on the mean cluster metallicity. The
formation history of the cluster may be important too. Therefore, studying the
properties of the evolved stars in $\omega$ Cen can help us understand the
evolution of stars with $M_{\rm ZAMS}\sim0.8$ M$_\odot$ and the formation and
evolution of globular clusters, but it will probably not provide a complete
picture. Also, as the mean metallicity is not as high as in some other
globular clusters that have [Fe/H]$>-1$, it might be difficult to probe and
study dust in the outflows of red giants in $\omega$ Cen, and the effect of
this dust on the further evolution of the stars as well as its fate in the
intra-cluster medium.

\section{Scientific objectives}

With the availability of multi-fibre spectroscopes at large telescopes, and a
magnitude-limited inventory of membership and photometry for the upper part of
the Hertzsprung-Russell Diagram (HRD) of $\omega$ Cen, it was decided to
perform a spectroscopic survey of a significant fraction of the number of
post-Main Sequence stars. The aim was to obtain effective temperatures and
elemental abundances for stars across the upper HRD, as well as other
diagnostics for the conditions in the stellar atmospheres such as gravity,
micro-turbulence, rotation and temperature inversion layers, which in
principle will provide us with a map on which the path of evolution is
outlined by the changing structure of the stellar atmosphere and interior.
This can be used to test the results obtained from photometric studies, to
break degeneracies encountered in photometric analyses, and to put previous
spectroscopic work into a global perspective in order to arrive at a
consistent picture of post-Main Sequence evolution in $\omega$ Cen. In
addition we may find spectroscopic binaries, measure radial velocities to
confirm membership and study the kinematics within the gravitational
potential, or detect faint absorption by the intra-cluster medium as well as
the intervening inter-stellar medium.

\subsection{Abundances}

The spread in metallicities as deduced from photometric and limited
spectroscopic studies of RGB stars needs to be confirmed and refined using
direct spectroscopic determinations of a statistically sound sample of stars.
Although ideally these would be Main Sequence stars with photospheres in Local
Thermodynamic Equilibrium, stars on the RGB are next best, having suffered
relatively little from astrophysical processes that enrich or deplete surface
abundances. Care should be taken to also sample the wings of the distribution
in order not to be biased towards the most common examples. This can then set
the stage for careful modelling of the stellar ages. The initial metallicity,
exact ZAMS mass and binarity are some of the most fundamental parameters that
will determine the further evolution of a star, and this should be consistent
with observations of stars that have evolved beyond the base of the RGB.
Therefore, the metallicity distribution should be investigated at different
heights along the RGB and at subsequent evolutionary stages such as the
Horizontal Branch (HB) and Asymptotic Giant Branch (AGB). More detailed
analysis of the abundance spectrum may reveal the effects of astrophysical
processes such as mixing, magnetic fields, nuclear burning and stellar winds
on the post-Main Sequence evolution.

\subsection{Variability}

Many variable stars are known in $\omega$ Cen (Kaluzny et al.\ 1997; van
Leeuwen et al.\ 2000), most of which are radial pulsators. Pulsation of
stellar mantles is an important phenomenon because it affects the structure of
the stellar surface and interior, and it can affect the evolution of the star
through mixing and/or mass loss as well as the evolution of close binaries.
Pulsation can also be used as a tool to determine global stellar parameters
and/or to probe the structure of the stellar interior. For instance, the
pulsation period and luminosity can be used to estimate the (pulsation) mass
of the star, which can then be reconciled with the predicted mass on the basis
of evolutionary models. The observed discrepancy between these two estimates
for Cepheid variables is often interpreted as evidence for the loss of
${\Delta}M\sim0.2$ M$_\odot$ on the RGB.

The number of variable stars is dominated by RR Lyrae variables in the
instability strip between the RGB and blue-HB, with typical periods around
half a day. In $\omega$ Cen the range in $(B-V)$ colours of the RR Lyrae stars
is surprisingly broad. This might be due to an extended range in $T_{\rm eff}$
for stars in the instability strip as a result of the metallicity spread in
$\omega$ Cen. This could then also be responsible for the observed large
number of RR Lyrae colour ``mimics'' --- stars that have similar $(B-V)$
colours and luminosities as the RR Lyrae variables but that do {\em not}
pulsate. This hypothesis may be tested spectroscopically.

Other variables include bright red Long Period Variables (LPVs) with typical
periods of (a few) 100 days. These are commonly found on the AGB, but there
are indications that stars near the tip of the RGB also pulsate. Spectroscopy
of these LPVs may shed light on their RGB or AGB nature, on the particular
conditions in the stellar mantles and the possible connection with significant
mass loss. A few more blue variables are found at luminosities above the HB
but below the evolutionary tracks of post-AGB stars, and their nature is yet
to be determined (see van Leeuwen et al.\ 2000). They may be on their way from
the blue-HB to the early AGB, or they may be stars that do not make it all the
way onto the AGB --- the so-called ``AGB manqu\'{e}'' stars. Again,
spectroscopic analysis of the content and conditions in their photospheres may
provide an answer.

\section{Observations}

\begin{figure}
\plotfiddle{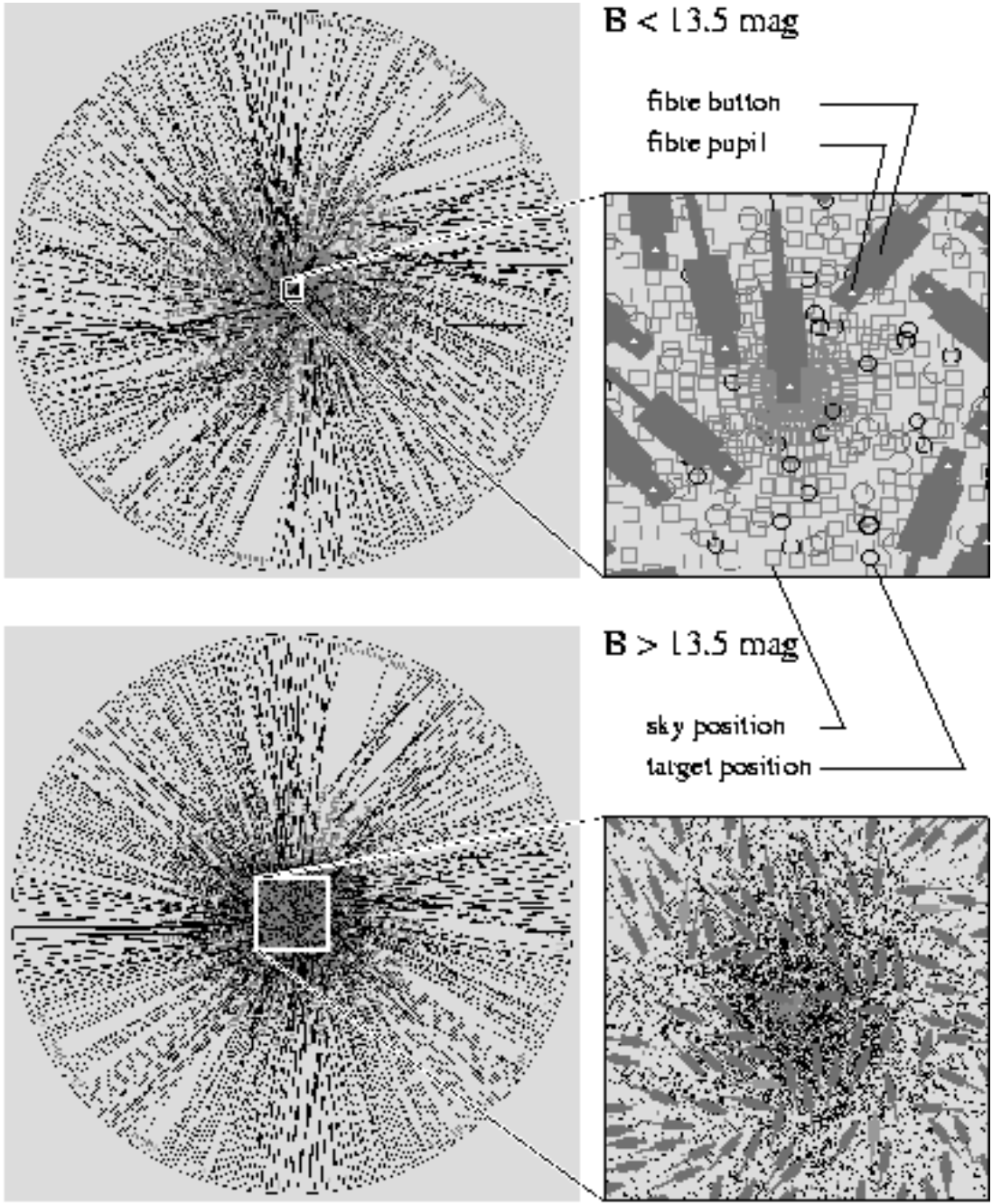}{181mm}{0}{100}{100}{-308mm}{-125}
\caption{Examples of fibre configurations used in our observations: for the
brighter (top) and fainter (bottom) targets. The left handside shows the
entire 2 degree field, and the right handside shows a close-up on the central
region where most of the targets (circles) are. The fibre pupil is much
smaller (0.14 mm) than the fibre button ($2\times5$ mm$^2$).}
\end{figure}

Multi-object spectroscopy was performed using the 2dF (two degree field)
fibre-fed spectroscope at the 4m Anglo Australian Telescope on the four half
nights of 26 \& 27 February and 1 \& 2 March 2000. The 2dF instrument provides
400 fibres of $2.1^{\prime\prime}$ for the acquisition of light from stars or
sky positioned within a field diameter of $2^\circ$. Two (thinned) CCDs of
1k$\times$1k each record 200 spectra produced by the fibre-fed spectrograph.
Four additional fibres are used for field acquisition and guiding on four
bright stars (fiducials). Fig.\ 1 shows typical fibre allocations for stars in
$\omega$ Cen. The extent to which stars in the core of the cluster can be
accessed is not so much due to crowding relative to the fibre aperture but due
to the considerable physical size of the button which holds the fibre. Grating
1200B was chosen to acquire spectra from $\lambda=3835$ up to $4965$ \AA\
(B-band) at a spectral resolution of $\Delta\lambda\sim2$ \AA\ ($R\sim2000$)
with an efficiency of 50 to 60\%. The sample of targets was divided into a
sample of bright targets with $B<13.5$ mag, which spectra were integrated for
$3\times5$ minutes, and a sample of faint targets with $B\geq13.5$ mag, which
spectra were integrated for $3\times20$ minutes. While one plate of fibre
allocations was being exposed, another plate of fibres was being prepared for
the next set of targets.

The data were reduced within the dedicated 2dFdr software package. This
includes bias subtraction and flatfield correction, wavelength calibration
using spectra of an arc lamp (Fe/Ar), spectrum tracing and extraction, and a
correction for scattered light. Sky subtraction was performed offline by
constructing a radial grid of sky spectra in order to account for the
contribution to the sky from the unresolved stellar population of $\omega$
Cen. The contamination within the fibre aperture by the light from
neighbouring stars was estimated by using the photometric survey data from
Kaluzny et al.\ (1997), which is fairly complete down to the limits of our
spectroscopic survey. Only 15 (1\%) of the observed stars and 4 ($<1$\%) of
the observed sky positions suffer from contamination in excess of 1\%, with a
maximum of $\sim40$\%.

\begin{figure}
\plotfiddle{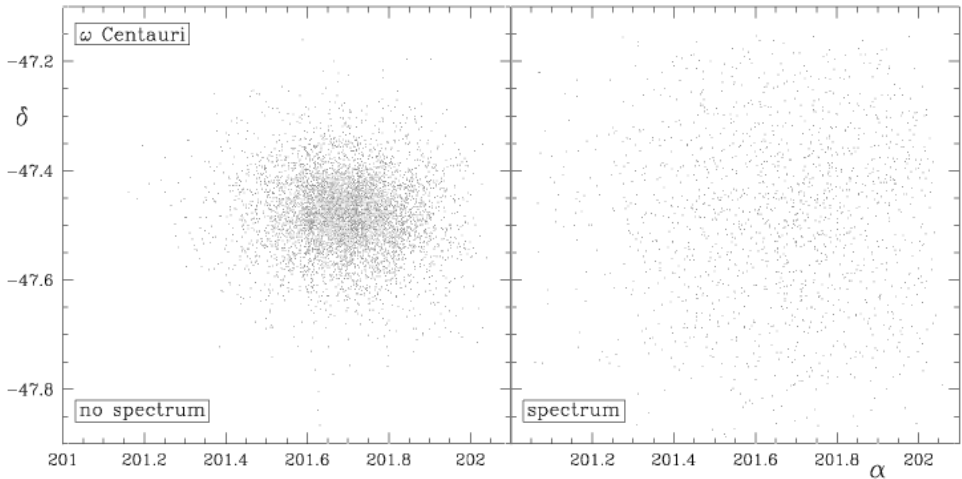}{55mm}{0}{50}{50}{-160mm}{-120mm}
\caption{Spatial distribution on the sky (in degrees) of members of $\omega$
Cen for which spectra were (right) or were not (left) obtained.}
\end{figure}

\section{Target selection}

\begin{figure}
\plotfiddle{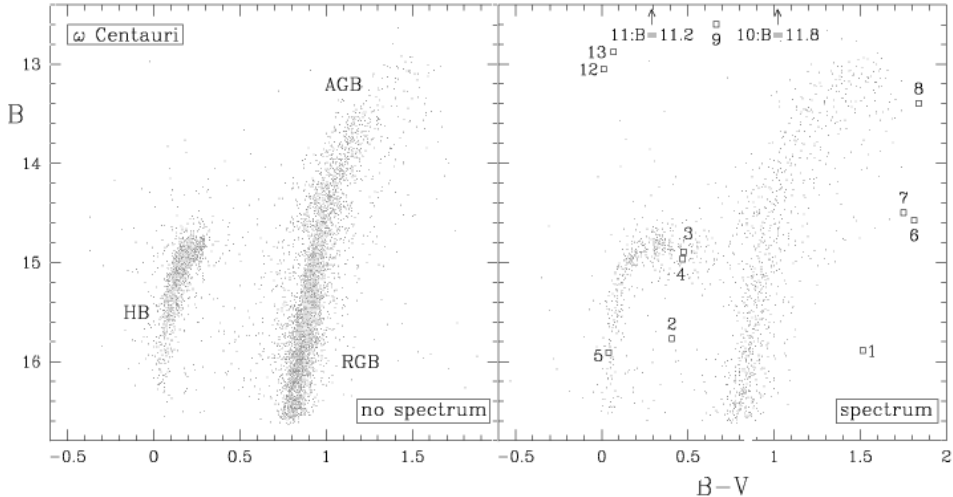}{60mm}{0}{50}{50}{-156mm}{-113mm}
\caption{Distribution across the Hertzsprung-Russell Diagram of members of
$\omega$ Cen for which spectra were (right) or were not (left) obtained. The
examples described here are labelled from 1 to 13.}
\end{figure}

The main objective was to cover the upper HRD in a uniform fashion, as much as
possible including outliers. Initially, the proper motion survey of van
Leeuwen et al. (2000) provided us with 9847 objects, of which we selected
stars with a $p\geq90$\% probability of cluster membership. Stars were
rejected if they had other known objects within a distance of
${\Delta}<2^{\prime\prime}$ or (in a few cases only) if no $B$ \& $(B-V)$
photometry was available for them. The uniform colour-magnitude selection was
established on the basis of a grid with ${\Delta}B=0.1$ \&
${\Delta}(B-V)=0.05$ mag. For each of these grid boxes, stars were assigned
priorities according to their distance from the cluster centre, with the
highest priority for the most distant star --- which was thought to probably
yield the highest quality spectra for stars of a particular magnitude and
colour. Some boxes contain as few as zero objects, but from none of the boxes
were drawn more than a few objects. In addition to the uniformly selected
sample, the interest in the RR Lyrae variability strip led us to include all
150 variables from van Leeuwen et al. (2000) and all 134 RR Lyrae colour
mimics with $0.3<(B-V)<0.6$ \& $14.5<B<15.2$ mag, plus the well-known and
brightest cluster member HD 116745 (Fehrenbach's star). The remaining
brightest red giants were used to define a sample of cluster members to act as
fiducials. The sky positions were drawn from a concentric grid of 2295
positions at an increasing density towards the cluster core after excluding
positions with a known star within $2^{\prime\prime}$.

A total of 9 configurations were used to obtain spectra of 1528 different
stars in $\omega$ Cen, of which 238 stars were repeated on another occasion,
down to $B\sim16.6$ mag at S/N$\sim40$, together with a total of 1135 spectra
on 692 unique sky positions. The distributions on the sky and across the HRD
of the $p\geq90$\% cluster members and the observed stars amongst them are
shown in Figs.\ 2 \& 3, respectively. As a result of the selection procedure
the spatial distribution is not centrally concentrated. The HRD is covered
relatively uniformly except for the RR Lyrae region from which most targets
were in fact observed. However, quite many of the stars from sparsely
populated regions in the HRD could not be observed because they tend to lie in
the central densest regions of the cluster.

\section{Examples}

To give a flavour of the quality of the spectra and their potential use, a few
examples are presented and briefly discussed. Labelled from 1 to 13, they are
more or less ordered along the evolutionary path of a star (Fig.\ 3) as it
ascends the RGB, progresses through the HB and ends via the AGB as a post-AGB
object. The 5-digit numbers in the spectra are their entries in the proper
motion catalogue of van Leeuwen et al.\ (2000). All spectra are clearly
red-shifted by $\Delta\lambda\sim3$ \AA\ with respect to the laboratory
wavelengths of the strongest lines, which is consistent with membership of
$\omega$ Cen ($v_{\rm rad}\sim230$ km s$^{-1}$).

\subsection{Red and blue RGB stars}

\begin{figure}
\plotfiddle{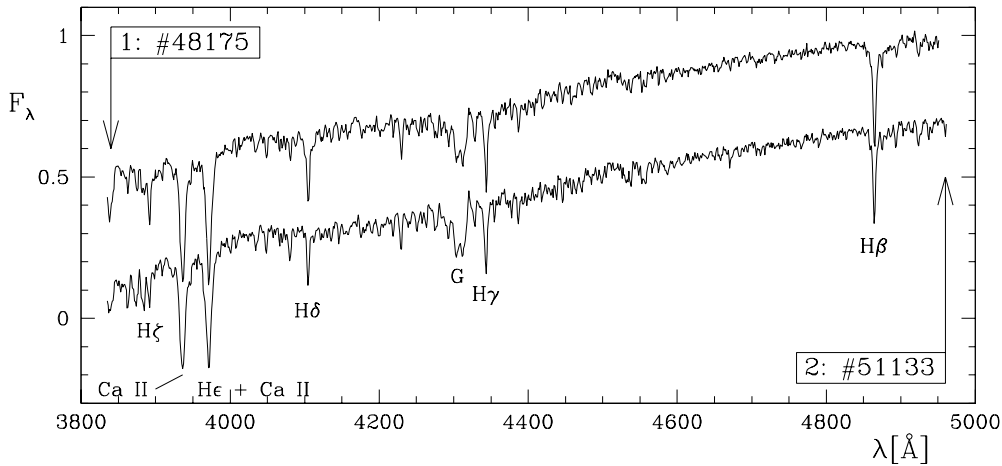}{55mm}{0}{70}{70}{-215mm}{-128mm}
\caption{Two stars with $B$ \& $V$ magnitudes like very red (\#1) and very
blue (\#2, offset by $-0.3$) RGB stars.}
\end{figure}

After having left the Main Sequence, 0.8 M$_\odot$ stars spend most of their
remaining life on the RGB. Therefore, the RGB is densely populated and ideal
for the study of the metallicity distribution. Photometric determinations of
the mean value for the metallicity are often quite accurate, but photometry in
the crowded core of globular clusters causes observational scatter which might
mimic extended wings in the metallicity distribution. As a cautionary example,
therefore, two stars were selected that have $B$ \& $V$ magnitudes ressembling
respectively very red and very blue RGB stars. Their K-type spectra turn out
to be virtually identical (Fig.\ 4): weak hydrogen Balmer lines, but strong
absorption by the Ca {\sc ii} H \& K lines and by the G-band. Many weak
absorption lines from metals (i.e.\ not noise!) can be seen, and these may be
used in future analysis of the detailed abundance spectrum of the RGB stars.
These two stars are only a few arcminutes away from the centre of $\omega$
Cen, and hence their photometry is less reliable. The red star is the only one
amongst the 12 stars with $B>15.5$ \& $(B-V)>1.2$ mag for which a spectrum was
obtained, because they all lie in the core region of $\omega$ Cen.

\subsection{RR Lyrae variables and colour mimics}

The RGB evolution terminates with the ignition of helium in the stellar core.
If the star is of low mass and metallicity, it will continue to evolve along
the HB, passing through a phase where the mantle is unstable against radial
pulsation. These pulsations are easily detected as photometric variability
with amplitudes of tenths of a magnitude and periods of several hours up to a
day. An example of the spectrum of such an RR Lyrae-type variable ($P=0.79$
days) in $\omega$ Cen is presented in Fig.\ 5, along with the spectrum of
another member of $\omega$ Cen which has almost identical $B$ \& $V$
magnitudes but which is not seen to pulsate --- i.e.\ a colour mimic. The
spectrum of the variable star contains many faint absorption lines of metals
and strong absorption by Ca {\sc ii}, in stark contrast to the spectrum of the
non-variable star that is dominated by hydrogen Balmer lines with remarkably
strong absorption wings. A more detailed analysis of the chemical content and
physical conditions of the photospheres of these and other RR Lyrae variables
and colour mimics may clarify the reason for both their differences and
similarities. Also, for some of these stars, spectra were taken more than once
and hence their variability may be studied spectroscopically.

\begin{figure}
\plotfiddle{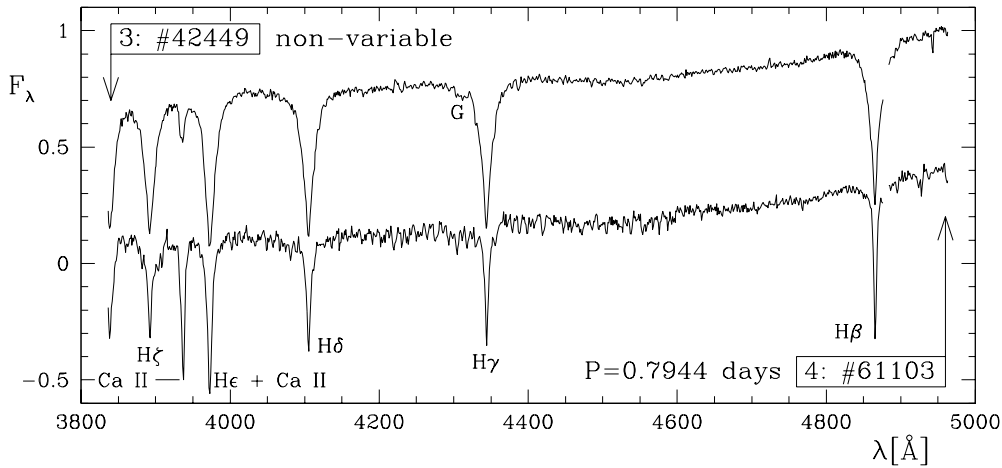}{55mm}{0}{70}{70}{-215mm}{-128mm}
\caption{A variable star of RR Lyrae type (\#4, offset by $-0.6$) and a
non-variable RR Lyrae colour mimic (\#3).}
\end{figure}

\subsection{Extreme horizontal branch}

In some globular clusters, the HB extends to very blue colours indicative of
high photospheric temperatures. In extreme cases such as observed in $\omega$
Cen, the photosphere can become so hot that the maximum of continuum emission
shifts sufficiently into the ultra-violet to render optical photometry
insensitive to $T_{\rm eff}$. In the $B$ versus $(B-V)$ diagram, for instance
(Fig.\ 3), the extreme blue-HB bends over to become a ``Vertical Branch'' near
$(B-V)\sim0$ mag with the stars becoming optically fainter as more of their
light is radiated at wavelengths $\lambda<3000$ \AA. The morphology of the HB
is observed to differ substantially amongst globular clusters with similar
metallicity, and it is not yet understood what is the ``second parameter(s)''
responsible for this: age in combination with mass loss on the RGB may not be
sufficient (Catelan 2000), but binary evolution may be important (Maxted et
al.\ 2001). An example of an extreme blue-HB star is given in Fig.\ 6: a hot
continuum with strong hydrogen Balmer absorption lines, very weak He {\sc i}
and no Ca {\sc ii} (except for an interstellar component). As this is one of
the fainter stars in our spectroscopic sample, most of the structure seen in
the continuum is probably due to noise.

\subsection{Cool giants}

After core-helium burning extinguishes, stars of $M_{\rm ZAMS}\sim0.8$
M$_\odot$ may ignite hydrogen in a shell surrounding the electron-degenerate
core and ascend the AGB. Stars that fail to do so (AGB manqu\'{e} stars) drop
off the HB towards the white dwarf cooling tracks. In some colour-magnitude
diagrams the early-AGB blends with the RGB. In the $B$ versus $(B-V)$ diagram
of $\omega$ Cen (Fig.\ 3), the RGB-tip is reached at $B\sim13.5$ mag and hence
the (red) stars brighter than this must be on the AGB. However, as the mantle
of both RGB and AGB stars inflates enormously, their photosphere cools to
$T_{\rm eff}<4000$ K, radiating most of the light at wavelengths
$\lambda>7000$ \AA. Hence their optical brightness diminishes if their colours
become redder than $(B-V)>1.5$ mag. If circumstellar dust is formed around
these cool stars, they become even redder and fainter at optical wavelengths.
Because the mantles of very cool giants become unstable against radial
pulsations, and because the surface gravity is very low and the formation of
molecules and/or dust in their extended atmospheres promotes driving a wind,
AGB stars enter a phase of high mass-loss rates ($\dot{M}>10^{-7}$ M$_\odot$
yr$^{-1}$) during which they shed their entire mantles (van Loon 2001).
What happens for stars near the tip of the RGB is not yet known, but it is
believed that as much as ${\Delta}M\sim0.2$ M$_\odot$ much be lost on the RGB
(Fusi Pecci \& Renzini 1976).

\begin{figure}
\plotfiddle{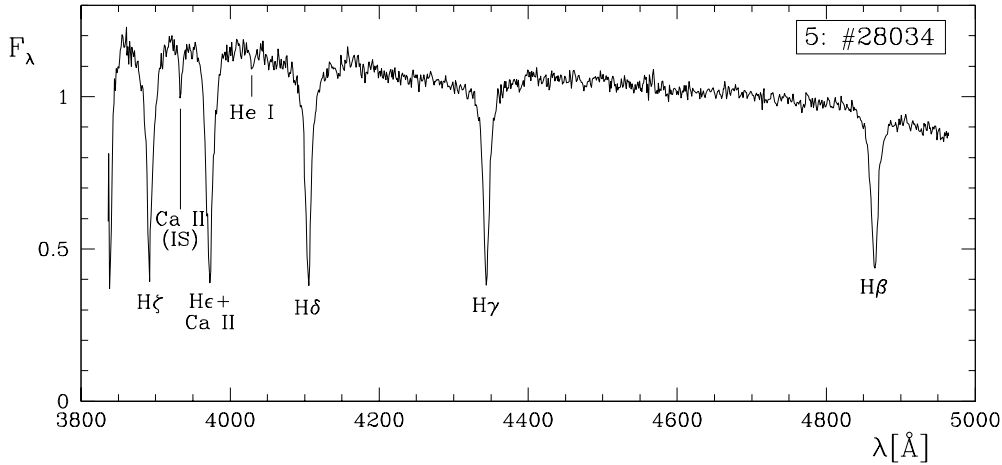}{55mm}{0}{70}{70}{-215mm}{-128mm}
\caption{An example of an extreme blue-HB star.}
\end{figure}

The spectra of two very cool stars (M3--4) in $\omega$ Cen are presented in
Fig.\ 7. It is not certain whether these stars are on the AGB or on the RGB,
although their position in the $B$ versus $(B-V)$ diagram (Fig.\ 3) suggests
they may be at the tip of the distinct red RGB that is currently being
interpreted as due to a (small) population of relatively metal-rich stars
(Pancino et al.\ 2001, and references therein). Their photospheres are
sufficiently cool to show strong absorption by TiO molecules. One of the two
stars exhibits hydrogen Balmer line emission which may arise from shocks
travelling through a pulsating atmosphere. Indeed, this star, \#33062 is known
to be variable in an irregular way, but the timescale of variability of
$P\sim0.5$ days (van Leeuwen et al.\ 2000) is surprisingly short for a cool
giant and inconsistent with radial pulsation. The other star, \#35250 is also
a known irregular variable with $P\sim51.3$ days --- typical for pulsation of
a red giant even though its spectrum does not show line emission.

\begin{figure}
\plotfiddle{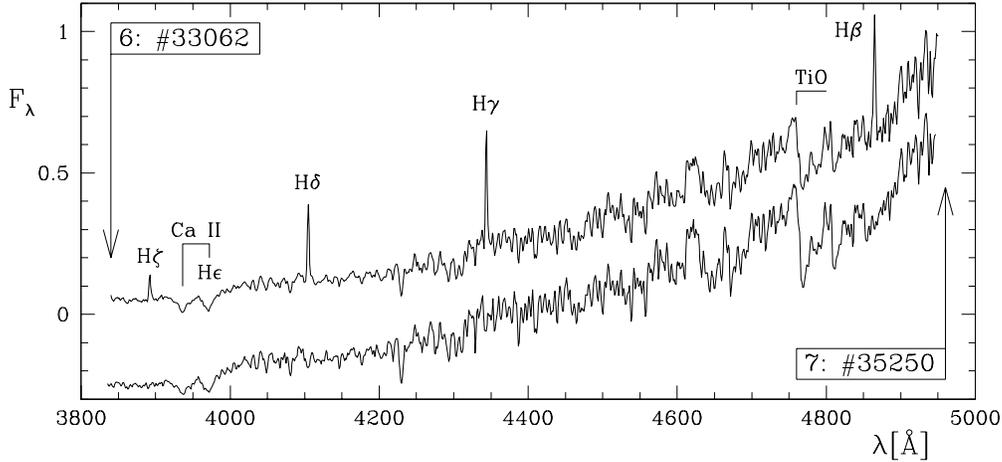}{55mm}{0}{70}{70}{-215mm}{-128mm}
\caption{Two very cool RGB or AGB stars with (\#6) and without (\#7, offset by
$-0.3$) hydrogen Balmer line emission.}
\end{figure}

\subsection{A reddened AGB star?}

\begin{figure}
\plotfiddle{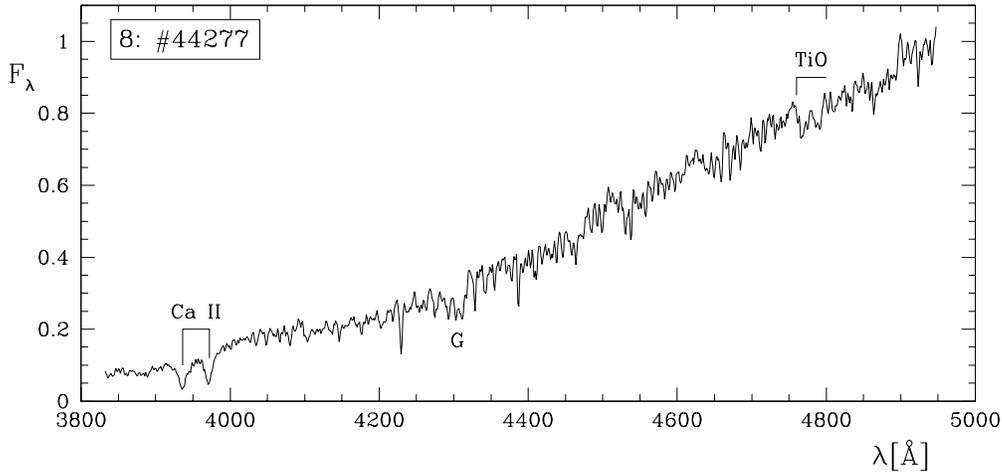}{55mm}{0}{70}{70}{-215mm}{-128mm}
\caption{This AGB star is surprisingly red for its spectral type.}
\end{figure}

With $(B-V)=1.8$ mag, \#44277 is the reddest star of which a spectrum was
obtained (Fig.\ 8), and indeed the reddest member of $\omega$ Cen down to
$B\sim16$ mag. It is probably an AGB star of spectral type about M0. The
spectral slope, however, is much steeper than expected for such spectral type:
compare, for instance, with the spectra of the much cooler giants in Fig.\ 7.
Is the photosphere of this star really as cool as the spectral slope suggests,
and does the early spectral type merely reflect a very low metallicity? Or is
the spectral slope reddened by intervening dust? The latter explanation is
quite plausible for a star near the tip of the AGB: the dust may be of
circumstellar origin and have formed in the outflow from the AGB star. The
rather moderate spectral type may indeed be due to low metallicity, causing a
consequently warm photosphere for a mass-losing AGB star and hence a
relatively small stellar radius. The (irregular) variability and its
relatively short period of $P\sim113$ days (van Leeuwen et al.\ 2000) are
consistent with the interpretation of a metal-poor mass-losing AGB star. The
mass-loss rate must then be quite high in order to produce as much dust as is
needed to cause the observed reddening, as the dust-to-gas ratio in the
outflow is (probably linearly) proportional to the metallicity (van Loon
2001).

\subsection{Luminous post-AGB stars}

\begin{figure}
\plotfiddle{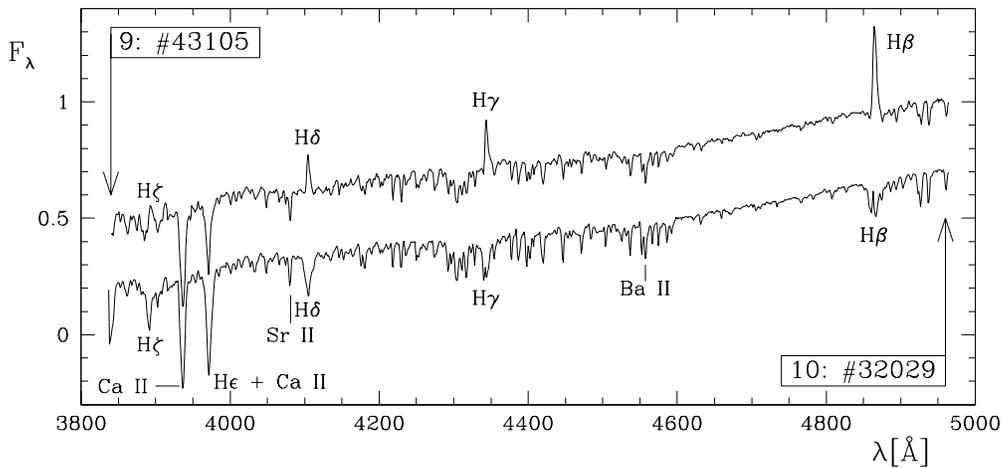}{55mm}{0}{70}{70}{-215mm}{-128mm}
\caption{Two bright post-AGB stars with hydrogen Balmer line emission (\#9) or
partly filled-in absorption (\#10, offset by $-0.3$).}
\end{figure}

The evolution along the AGB terminates when most of the mantle has been
consumed by nuclear burning and/or has disappeared as a result of mass loss.
Mass loss dominates over mass consumption for AGB stars of $M_{\rm ZAMS}>1$
M$_\odot$ (van Loon 2001), but this is not well documented for less massive
stars. The star leaves the AGB because the hot core becomes exposed and hence
the photospheric temperature increases dramatically, at first whilst
maintaining its high luminosity before the nuclear-burning shell finally
extinguishes. By then, the star has become a white dwarf that gradually cools
and faints. Two examples of such luminous post-AGB stars in $\omega$ Cen are
presented in Fig.\ 9. Their spectra show hydrogen Balmer line emission in one
case (\#43105), and partly filled-in absorption in the other case (\#32029)
where the contrast with the bright continuum is smaller. The line emission is
probably of circumstellar origin, the hot core illuminating the rarefied
material from its former mantle. Both spectra show quite strong Sr {\sc ii}
and Ba {\sc ii} absorption lines indicative of an enhancement of s-process
material which was dredged-up during the AGB phase from the nuclear burning
layers through the convective mantle into the photosphere (van Loon 2001).
Both stars are known variables with periods of $P=29$ days for \#43105 and
$P=32$ days for \#32029 (van Leeuwen et al.\ 2000).

\subsection{Fehrenbach's star}

\begin{figure}
\plotfiddle{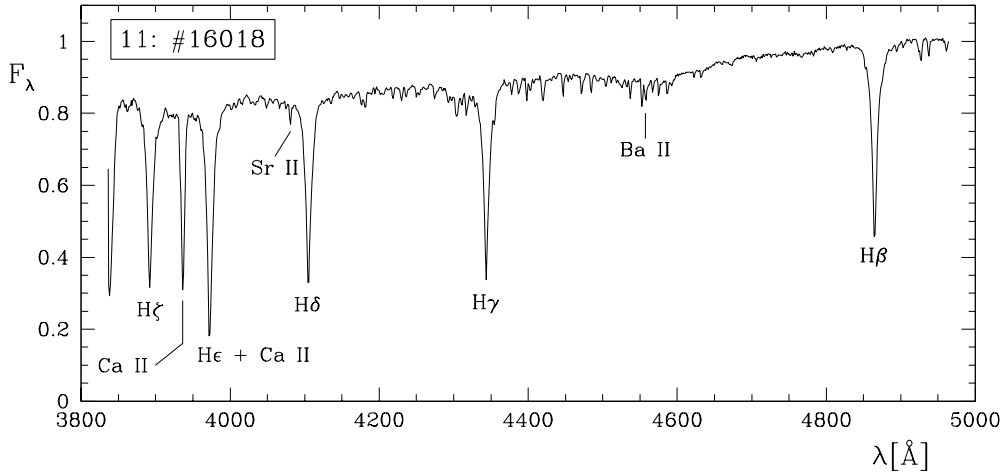}{55mm}{0}{70}{70}{-215mm}{-128mm}
\caption{The bright post-AGB star HD 116745 (Fehrenbach's star).}
\end{figure}

The brightest member of $\omega$ Cen in the B-band, Fehrenbach's star (HD
116745) is a post-AGB star (Gonzalez \& Wallerstein 1992). Its spectrum (Fig.\
10) shows strong hydrogen Balmer lines in absorption: line emission may be too
weak to see against the bright continuum, or alternatively the circumstellar
ionized material may already have dispersed by the time it took the star to
evolve to its present position in the HRD. Again, absorption lines of the
s-process elements Sr {\sc ii} and Ba {\sc ii} are indicative of the post-AGB
nature of this object.

\subsection{Luminous hot stars}

\begin{figure}
\plotfiddle{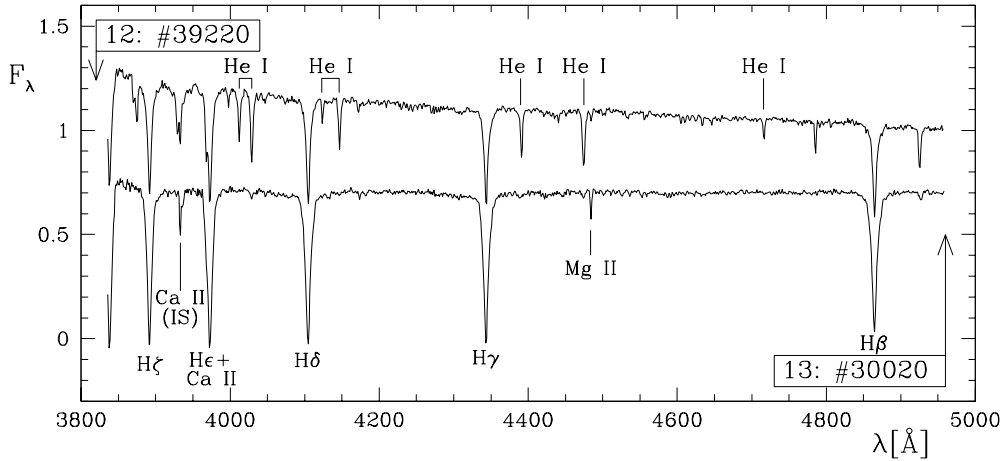}{55mm}{0}{70}{70}{-215mm}{-128mm}
\caption{Two bright hot stars with strong He {\sc i} lines (\#12) or an almost
purely hydrogen spectrum (\#13, offset by $-0.3$).}
\end{figure}

As a last example, spectra are presented of two bright hot stars (Fig.\ 11),
which presumably are post-AGB objects. Despite them having very similar B \&
V-band photometry, their spectra are quite different. Both have hot continua,
but star \#39220 is clearly (much) hotter than star \#30020 as evidenced by
the violet upturn of the continuum and the strong He {\sc i} absorption lines.
There is only a weak trace of the strongest amongst the He {\sc i} lines in
the spectrum of \#30020 which, besides strong hydrogen Balmer absorption
lines, is almost featureless except for an interstellar component of Ca {\sc
ii} and a narrow (interstellar?) Mg {\sc ii} absorption line. The hydrogen
Balmer lines in the spectrum of \#39220 are relatively weak, possibly due to
the high $T_{\rm eff}$ and/or a hydrogen deficiency in the photosphere of this
highly evolved post-AGB object.

\section{Status and future}

Using the 2dF instrument at the AAT, we have obtained 1766 high quality
spectra of 1528 confirmed members of $\omega$ Cen, in the wavelength range of
$\lambda=3835$ to $4965$ \AA\ at a resolving power of $R\sim2000$. It covers
the upper regions of the HRD including almost completely the HB and most of
the RGB, with the emphasis on sampling the extremes of the distribution of
stars in the HRD plus the variable and non-variable stars in the RR Lyrae part
of the instability strip.

The data reduction is nearly finalised, and all spectra will eventually be
made available to the astronomical community. The analysis of the spectra is
about to start, and will first aim at spectral classification in order to
construct a physical HRD ($L, T_{\rm eff}$). When possible, metallicities and
abundance spectra will be derived, which can then be used to reconstruct the
star formation and chemical enrichment histories of $\omega$ Cen. The changes
in the abundance spectra along the evolutionary paths in the HRD may help us
understand the morphology of the HB, the properties of the RR Lyrae variables
and RR Lyrae colour mimics, and dredge-up processes on the RGB and AGB.
Emission-line objects, reddened stars and post-AGB candidates will receive
particular attention in order to study mass loss from red giants in globular
clusters.

Future multi-object spectroscopes with smaller fibre pupils, finer gratings,
larger CCDs, at 8m-class telescopes allow the described spectroscopic study to
be extended: (i) in wavelength, especially 5000 to 9000 \AA, to cover more
spectral diagnostics for the determination of $T_{\rm eff}$, the abundance
spectrum, surface gravity, etcetera; (ii) to higher spectral resolution in
order to better measure line profiles for analysis of the physical conditions
in the photosphere, decompose blended lines and measure kinematics; (iii) to
fainter stars, including main-sequence stars (see the contribution by Russell
Cannon to these proceedings); (iv) to other galactic globular clusters and
nearby dwarf spheroidals.

\acknowledgments

I would like to thank the organisers for giving me the opportunity to present
this work at a pleasant and interesting conference, all the participants from
whom I have learnt a lot, and Joana.

\end{document}